# Altitude effects of localized source currents on magnetotelluric responses

Shinya Sato

Graduate School of Engineering, Kyoto University

Email: sato.shinya.27s@st.kyoto-u.ac.jp

**Abstract**

The effects of localized source currents on Earth's magnetotelluric (MT) responses have been reported in the literature in terms of the changes in period and subsurface structure. The focus in this study is on the bias within the MT responses arising from variations in the vertical and horizontal distances of the source current. The MT responses at 20 and 200 s were calculated at various distances from the source current. A slight change in source distance causes a shift in the MT responses, and the bias is large, especially over the altitudes explored in the MT data analysis (i.e., 100–150 km), where the E layer exists. The vertical distance of the source field varies because the distribution of conductivity with altitude in the ionosphere and the region controlling the ionospheric electrical process change temporally. Thus, in assessing the temporal changes in MT responses, we should treat them carefully by checking the ionospheric environment.

**Introduction**

In magnetotelluric (MT) surveys, the primary source magnetic fields are assumed horizontally uniform. The effects of localized source currents on the MT responses have been discussed in the literature (Madden and Nelson, 1964; Schmucker, 1970; Hermance and Peltier, 1970; Häkkinen et al., 1989; Pirjola, 1992; Viljanen, 2012), where impedances over long periods and at sites above structures of high resistivity are biased because of the collapse of the plane-wave assumption. For example, Pirjola (1992) calculated the downward/upward bias of the apparent resistivities in the range of 1–100,000 s due to the electrojet with field–aligned currents under the condition that the divergence of the total current density vanishes. The study reported that the apparent resistivity of 100 Ωm and at periods larger than 60 s was clearly affected. Although past studies focused on period-dependent bias, the bias stemming from the variation in distances, especially altitudes, of the localized source currents was not discussed in detail.

Temporal/seasonal changes (Brändlein et al., 2012; Romano et al., 2014; Vargas and Ritter, 2016) and bias (Murphy and Egbert, 2018) in the MT responses/vertical geomagnetic transfer-functions due to the source field have been recently reported. For example, Romano et al. (2014) reported that, for time periods 20–100 s, the apparent resistivities have a negative correlation with geomagnetic activity. In particular, during time-lapse MT soundings, the source bias on the MT responses should be evaluated because we possibly misinterpret the temporal shifts in impedances of the source field as reflecting changes in the subsurface resistivity.

In this study, the electromagnetic fields and MT responses were calculated by varying the vertical and horizontal distances of the source current. The study revealed: i) the numerical examples of the bias in the MT responses because of the variation in vertical and horizontal distances of the source field, ii) implications from these examples, iii) the mathematical underpinning of this bias, and iv) the mathematical conditions for upholding the plane-wave assumption.

**Electromagnetic fields above Earth's surface**

We chose a Cartesian coordinate system, where the $x$, $y$, and $z$ axes are northward, westward, and downward positive, respectively, with $z = 0$ at Earth's surface. The coordinate system, the line source current to be defined later, and an observation site are depicted in Fig. 1.

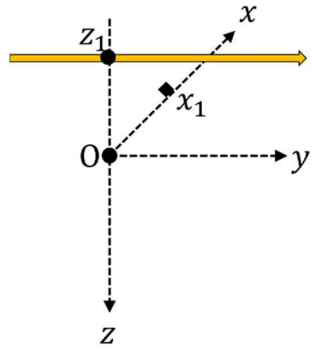

Figure 1. Coordinate system used in this study. The symbol of "O" and diamond symbols mark the origin and an observation site, respectively; $z_1$ and $x_1$ are the altitude of the line source current and the horizontal distance of an observed site, respectively.

Ignoring the displacement current and using the SI system, Maxwell's equations in the frequency domain are

$$\nabla \times \mathbf{E} = -i\omega \mathbf{B}, \tag{1}$$
$$\nabla \times \mathbf{B} = \mu_0 (\sigma \mathbf{E} + \mathbf{J}), \tag{2}$$
$$\nabla \cdot \mathbf{B} = 0, \tag{3}$$

where $\mathbf{E}$, $\mathbf{B}$, and $\mathbf{J}$ are the electric field, magnetic induction, and source current, respectively; $\sigma$, $\mu_0$, and $\omega$ are the electrical conductivity, the magnetic permeability of free space, and the angular frequency, respectively. Introducing the vector potential $\mathbf{A}$ and the scalar potential $\Pi$, the electromagnetic fields are

$$\mathbf{B} = \nabla \times \mathbf{A}, \tag{4}$$
$$\mathbf{E} = -i\omega (\mathbf{A} + \nabla \Pi). \tag{5}$$

We apply a gauge transformation such that **A** and $\Pi$ satisfy

$$i\omega\sigma\mu_0\Pi = -\nabla \cdot \mathbf{A}, \tag{6}$$

and the equation for the vector potential can be represented as

$$-\Delta \mathbf{A} + i\omega\sigma\mu_0 \mathbf{A} = \mu_0 \mathbf{J}. \tag{7}$$

Because $\nabla \cdot \mathbf{E} = 0$, the equation for the scalar potential,

$$-\Delta\Pi + i\omega\sigma\mu_0\Pi = 0, \tag{8}$$

holds. Considering the electromagnetic fields above Earth's surface (i.e., $z \leq 0$), $\sigma$ may be taken as $\sigma_0$ denoting the electrical conductivity of free space.

As in the analysis in Hermance and Peltier (1970), we consider a wire at an altitude $z_1 < 0$ carrying an electric current $I$ (Fig. 1); the current density is

$$\mathbf{J} = \begin{pmatrix} J_x \\ J_y \\ J_z \end{pmatrix} = I\delta(z - z_1)\delta(x)\begin{pmatrix} 0 \\ 1 \\ 0 \end{pmatrix}. \tag{9}$$

We focus on only $A_y$, the $y$ component of **A**, and $\Pi$. For this study, the application of the Fourier transform (FT) to the horizontal components transforms $x$ and $y$ into the wavenumber domain yielding

$$\tilde{F}(\eta, \zeta) = \int_{-\infty}^{\infty}\int_{-\infty}^{\infty} F(x,y) e^{i(\eta x + \zeta y)} dx dy, \tag{10}$$

$$F(x,y) = \frac{1}{4\pi^2}\int_{-\infty}^{\infty}\int_{-\infty}^{\infty} \tilde{F}(\eta, \zeta) e^{-i(\eta x + \zeta y)} d\eta d\zeta. \tag{11}$$

Eq. 11 enables us to transform Eq. 7 into

$$\frac{\partial}{\partial z^2}\tilde{A}_y - \beta_0^2 \tilde{A}_y = -\mu_0 \tilde{J}_y \quad (z \leq 0), \tag{12}$$

where $\tilde{A}_y$ and $\tilde{J}_y$ are $A_y$ and $J_y$ in the wavenumber domain, respectively, and $\beta_0 = \sqrt{(\eta^2 + \zeta^2) + i\omega\mu_0\sigma_0}$. Eq. 12 is the Helmholtz equation and for which its Green's function satisfies

$$\frac{\partial}{\partial z^2} G(z, z') - \beta_0^2 G(z, z') = \delta(z - z'). \tag{13}$$

As shown in Arfken et al. (2012), the solution of $G(z, z')$ is

$$G(z, z') = -\frac{\left(e^{-\beta_0|z-z'|} + \Lambda e^{-\beta_0|z+z'|}\right)}{2\beta_0}, \tag{14}$$

where $\Lambda$ is a constant required to uphold the boundary condition at $z = 0$. Consider a structure beneath Earth's surface ($z > 0$) having a half-space of conductivity $\sigma_1$. The continuity of the electromagnetic fields parallel to the boundary yields

$$\Lambda(\eta, \zeta) = \frac{\beta_0 - \beta_1}{\beta_0 + \beta_1}, \tag{15}$$

where $\beta_1 = \sqrt{(\eta^2 + \zeta^2) + i\omega\mu_0\sigma_1}$. Applying the FT to the horizontal components, Eq. 10, $\tilde{J}_y$ in Eq. 12 becomes

$$\tilde{J}_y = 2\pi I \delta(z - z_1)\delta(\zeta). \tag{16}$$

Using Green's function, the solutions for $\tilde{A}_y$ are

$$\tilde{A}_y = \pi\mu_0 I \frac{(e^{-\beta_0|z-z_1|} + \Lambda e^{-\beta_0|z+z_1|})\delta(\zeta)}{\beta_0}. \tag{17}$$

Similarly, applying the inverse FT to the horizontal components, Eq. 11, and considering $\sigma_0 = 0$, $A_y$ is written as

$$A_y = \frac{\mu_0 I}{4\pi} \int_{-\infty}^{\infty} \frac{(e^{-|\eta|(z-z_1)} + \Lambda(\eta,0)e^{|\eta|(z+z_1)})}{|\eta|} e^{-i\eta x} d\eta \quad (z \leq 0). \tag{18}$$

The vector potential **A** has only a $y$-component and $A_y$ is independent of $y$. As a result, the divergence of **A** is equal to zero, and the scalar potential is ignored because both sides of Eq. 6 vanish. From Eqs. 4 and 5, the magnetic induction $B_x$ and the electric field $E_y$ of a site ($x = x_1$) at Earth's surface ($z = 0$) is written as

$$B_x = \frac{\mu_0 I}{2\pi} \int_{-\infty}^{\infty} \frac{\sqrt{\eta^2 + i\omega\mu_0\sigma_1}}{|\eta| + \sqrt{\eta^2 + i\omega\mu_0\sigma_1}} e^{|\eta|z_1} e^{-i\eta x_1} d\eta, \tag{19}$$

$$E_y = -i\omega \frac{\mu_0 I}{2\pi} \int_{-\infty}^{\infty} \frac{1}{|\eta| + \sqrt{\eta^2 + i\omega\mu_0\sigma_1}} e^{|\eta|z_1} e^{-i\eta x_1} d\eta. \tag{20}$$

Taking their ratio $E_y/B_x$ gives the impedance $Z_{yx}$

$$Z_{yx} = \frac{-i\omega \int_{-\infty}^{\infty} \frac{1}{|\eta| + \sqrt{\eta^2 + i\omega\mu_0\sigma_1}} e^{|\eta|z_1} e^{-i\eta x_1} d\eta}{\int_{-\infty}^{\infty} \frac{\sqrt{\eta^2 + i\omega\mu_0\sigma_1}}{|\eta| + \sqrt{\eta^2 + i\omega\mu_0\sigma_1}} e^{|\eta|z_1} e^{-i\eta x_1} d\eta}. \tag{21}$$

For this study, the apparent resistivity $\rho_{yx}$ is given by

$$\rho_{yx} = \frac{\mu_0}{2\pi f} |Z_{yx}|^2, \tag{22}$$

where $Z_{yx}$ is defined in Eq. 21. Note that hereafter, for simplicity, $x_1$ is defined as the horizontal distance to the source current; the distance unit used is "km" instead of "m" when stating horizontal/vertical distances $(x_1, z_1)$ although all the above values are calculated using the SI system of units.

**MT responses biased by source line current**

Here, the subsurface resistivity and the time period are set to 1000 Ωm (i.e., $\sigma_1 = 10^{-3}$ S/m) and 20 s, respectively. The subsurface resistivity has the same value as that used for the crust in Hermance and Peltier (1970).

By changing the altitude of the source current $z_1$ from –100 to –1000 km in increments of 5 km and the horizontal distances $x_1$ =1, 5, 10, 50, 100, 500, 1000, and 5000 km, the variation in the field components $B_x$ (Eq. 19) and $E_y$ (Eq. 20) as well as the impedance $Z_{yx}$ (Eq. 21) were determined. The integrals in Eqs. 19 and 20 were calculated using the discrete approximation; the convergence of each was verified. Note that the electric current $I$ in Eq. 9 is 1000 A although the value has no influence on $Z_{yx}$. The calculated values of the apparent resistivity (Eq. 22) and phase are shown in Fig. 2.

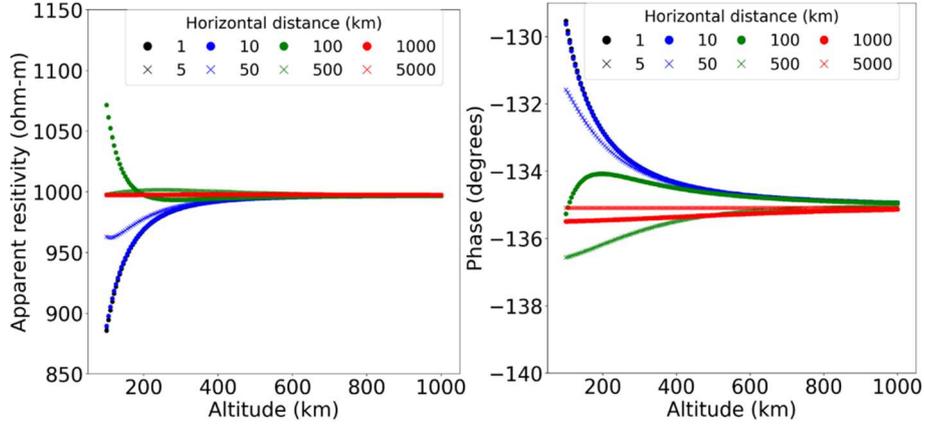

Figure 2. MT responses at a period of 20 s: (left) Relationship between the apparent resistivity and the altitude of the source current. (right) Relationship between the phase and the altitude of source current. The black/blue/green/red circles denote the responses derived from the horizontal distances of 1/10/100/1000 km, respectively. The black/blue/green/red crosses denote the responses derived from the horizontal distances of 5/50/500/5000 km, respectively.

In addition, the values of the apparent resistivity and phase are derived by varying $x_1$ from 0 to 3000 km in increments of 10 km and $z_1$ set to $z_1 =$ –100, –150, –200, –500, and –1000 km (Fig. 3).

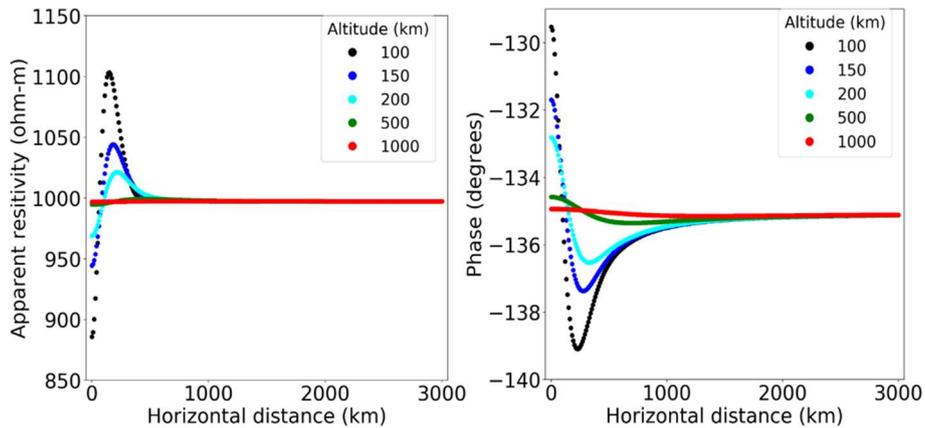

Figure 3. MT responses at a period of 20 s: (left) Relationship between the apparent resistivity and the horizontal distance of source current. (right) Relationship between the phase and the horizontal distance of source current. The black/blue/light blue/green/red circles denote the responses derived from the altitude of 100/150/200/500/1000 km, respectively.

On the basis of these results (Figs. 2 and 3), the MT responses are shifted largely depending on the vertical and horizontal distances from the source current, if the altitude is within the range 100–150 km where the E region (Viljanen, 2012; Sheng et al., 2014) with an important current system for MT exists. However, if $x_1$ and $|z_1|$ are larger, such bias becomes smaller.

The MT responses at a time period of 200 s under the same conditions as the above were plotted (Figs. 4 and 5) and, as expected, are biased by the source current more than those of Figs. 2 and 3.

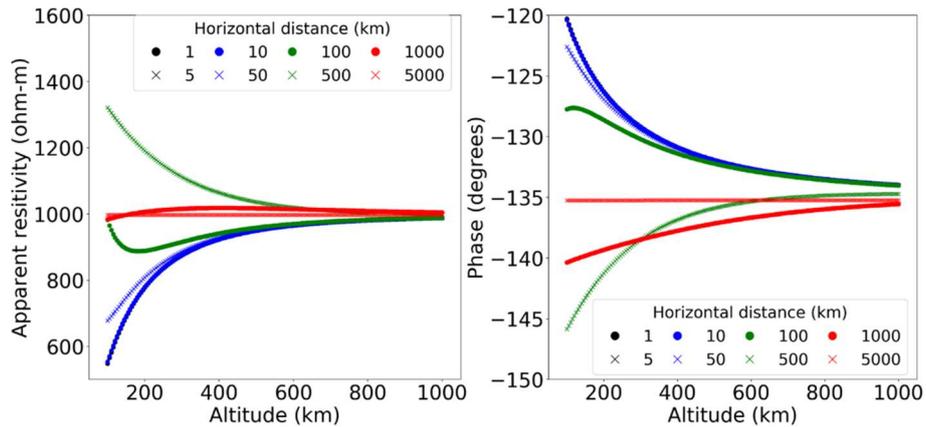

Figure 4. As for Fig. 2 except the period is set to 200 s.

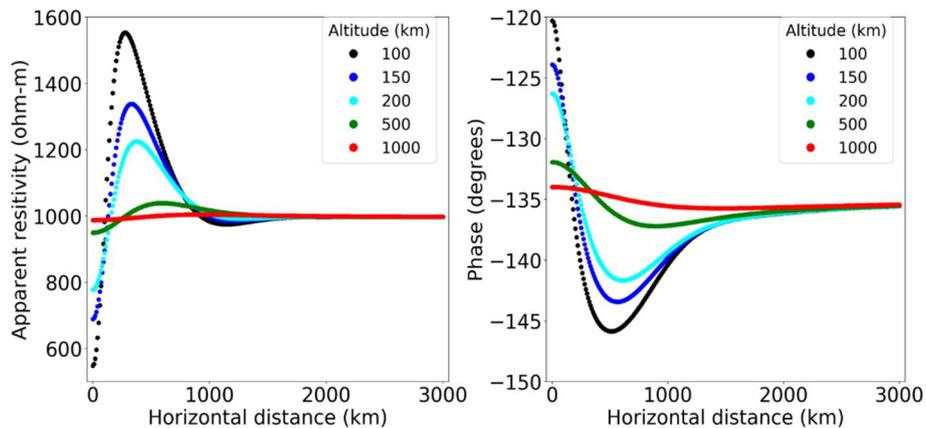

Figure 5. As for Fig. 3 except the period is set to 200 s.

Results from lower resistivity values (100/500 Ωm) are presented in Appendix A (Figs. S1–S8). These responses are less biased than the above cases (Figs. 2–5) as reported in the literature. If the conductivity and period are set to 0.01 S/m and 20 s, respectively, the bias within the MT responses approximately vanishes (Figs. S1 and S2). However, the other cases (Figs. S3–S8) are biased, especially if the altitude of the source current is within 100–150 km.

The line source current defined by Eq. 9 is possibly unrealistic, and maybe a sheet current is more suitable for actual current systems. The ionospheric current at mid-latitudes seems uniform over a wide area (Yamazaki and Maute, 2017). However, the current system for example at low latitudes is controlled by an equatorial electrojet (Yamazaki and Maute, 2017), which appears as a broad sheet current. As reported by McNish (1938), such current at altitudes higher than 100 km can be approximated by a line source current. Therefore, the entire analysis above, which has been derived assuming a line source current, is applicable to phenomena at low latitudes because in this study the altitude of the line current is at or above 100 km.

**Discussion**

A discussion is presented next of i) the mathematical basis of the bias on the MT responses due to the source field, ii) the mathematical condition upholding the plane-wave assumption, and iii) the implication arising from the numerical examples performed in this study.

The electromagnetic fields (Eqs. 19 and 20) generated by the line source current have an attenuation term

$$\alpha(\eta) = e^{|\eta|z_1}, \qquad (23)$$

and a term conveying information regarding the subsurface structure

$$\beta(\eta) = \sqrt{\eta^2 + i\omega\mu_0\sigma_1}. \qquad (24)$$

Substituting $\frac{2\pi}{20}$ 1/s, $1.26 \cdot 10^{-6}$ H/m, and 0.001 S/m for $\omega$, $\mu_0$, and $\sigma_1$, respectively, the apparent resistivity and phase were plotted (Figs. 2 and 3). When the wavenumber $|\eta|$ is greater than $2.0 \cdot 10^{-5}$, the influence of $\eta$ is greater than $\omega\mu_0\sigma_1$. If $|\eta|$ vanishes, $\alpha(\eta) = 1$ and the wavenumber effect on $\beta(\eta)$ vanishes. Hence, we set $\alpha(\eta) = 1$ for the standard. When $\alpha(\eta)$ in Eq. 23 is smaller than 0.01, the effect of $|\eta|$ is assumed negligible because the attenuation term $\alpha(\eta)$ is two orders of magnitude smaller than the standard (i.e., $\alpha(\eta) = 1$). To uphold this assumption, $|\eta|z_1$ should be smaller than $-4.6$, and when $z_1 = -100$ km, $|\eta|$ should be greater than $4.6 \cdot 10^{-5}$. Therefore, the integrands in Eqs. 19 and 20 are biased by wavenumber $|\eta|$ at least within the interval $2.0 \cdot 10^{-5} < |\eta| < 4.6 \cdot 10^{-5}$. However, given that $z_1 = -1000$ km, $\alpha(\eta) \leq 2.1 \cdot 10^{-9}$ when substituting $|\eta| \geq 2.0 \cdot 10^{-5}$, which yields a greater effect than $\omega\mu_0\sigma_1$. The wavenumber effect is small enough to be negligible, and as a result, the apparent resistivity approaches a constant value of 1000 Ωm, i.e., the subsurface resistivity. As expected from the above discussion, the MT responses at 200 s are biased more than those at 20 s because the influence of $\eta$ is greater than $\omega\mu_0\sigma_1$ when $|\eta|$ is greater than $6.3 \cdot 10^{-6}$. Moreover, on the basis of the above discussion, the MT responses at a site above a less resistive zone can be expected to be less biased (see Figs S1–S8 in Appendix A). For example, suppose the conductivity and period are set to 0.01 S/m (i.e., 100 Ωm) and 200 s, respectively, as in Figs. S5 and S6. Then, $\omega\mu_0\sigma_1$ in Eq. 24 calculated from these two variables has the same value as that obtained substituting 0.001 S/m and $\frac{2\pi}{20}$ 1/s (i.e., 20 s) into $\sigma_1$ and $\omega$. As a result, both MT responses at 200 s (Figs. S5 and S6) and at 20 s (Figs. 2 and 3) are shifted over a similar range (up to about 10%).

Both responses at 20 and 200 s indicate the same patterns of bias; that is, they shift depending not only on the vertical ($z_1$) but also on the horizontal distance ($x_1$) between the site and the source current. The integrands in Eqs. 19 and 20 oscillate because of the factor $e^{-i\eta x_1}$, and substituting 5000 km into $x_1$, the apparent resistivity (Eq. 22) at 20/200 s have values of 1000 $\Omega$m (see Figs. 2 and 4). The mathematics behind these calculated results resides with the re-expressions of Eqs. 19 and 20

$$B_x = \frac{\mu_0 I}{\pi} \int_0^\infty \frac{\sqrt{\eta^2+ia}}{\eta+\sqrt{\eta^2+ia}} \left( e^{\eta(z_1-ix_1)} + e^{\eta(z_1+ix_1)} \right) d\eta, \tag{25}$$

$$E_y = -i\omega \frac{\mu_0 I}{\pi} \int_0^\infty \frac{1}{\eta+\sqrt{\eta^2+ia}} \left( e^{\eta(z_1-ix_1)} + e^{\eta(z_1+ix_1)} \right) d\eta, \tag{26}$$

where $a = \omega \mu_0 \sigma_1$. Focusing on the term with $e^{\eta(z_1-ix_1)}$ and integrating by parts, we can obtain,

$$\int_0^\infty \frac{\sqrt{\eta^2+ia}}{\eta+\sqrt{\eta^2+ia}} e^{\eta(z_1-ix_1)} d\eta \tag{27}$$

$$= \frac{-1}{(z_1-ix_1)} + \frac{-1}{\sqrt{ia}(z_1-ix_1)^2} + \frac{-2}{ia(z_1-ix_1)^3} - \int_0^\infty \frac{d^3}{d\eta^3}\left(\frac{\sqrt{\eta^2+i}}{\eta+\sqrt{\eta^2+ia}}\right) \frac{e^{\eta(z_1-ix_1)}}{(z_1-ix_1)^3} d\eta,$$

$$\int_0^\infty \frac{e^{\eta(z_1-ix_1)}}{\eta+\sqrt{\eta^2+ia}} d\eta \tag{28}$$

$$= \frac{-1}{\sqrt{ia}(z_1-ix_1)} + \frac{-1}{ia(z_1-ix_1)^2} + \frac{-1}{(ia)^{\frac{3}{2}}(z_1-ix_1)^3} - \int_0^\infty \frac{d^3}{d\eta^3}\left(\frac{1}{\eta+\sqrt{\eta^2+ia}}\right) \frac{e^{\eta(z_1-ix_1)}}{(z_1-ix_1)^3} d\eta.$$

The triangle inequality and the inequality $\left|\frac{1}{(\eta^2+ia)}\right| \leq \frac{1}{a}$ ($\eta \in \mathbb{R}$) enable the integrands of the last terms in Eqs. 27 and 28 to be examined

$$\left|\frac{d^3}{d\eta^3}\left(\frac{\sqrt{\eta^2+ia}}{\eta+\sqrt{\eta^2+ia}}\right)\right| = \left|\frac{3(\eta-\sqrt{\eta^2+ia})}{(\eta+\sqrt{\eta^2+ia})(\eta^2+ia)^{\frac{3}{2}}}\left\{1 + \frac{\eta(\eta+2\sqrt{\eta^2+ia})}{(\eta^2+ia)}\right\}\right| \tag{29}$$

$$\leq \frac{6\eta+3a}{a^2}\left(1 + \frac{3\eta^2+2\sqrt{a}\eta}{a}\right),$$

$$\left|\frac{d^3}{d\eta^3}\left(\frac{1}{\eta+\sqrt{\eta^2+ia}}\right)\right| = \left|\frac{-3\eta}{(\eta^2+ai)^{\frac{5}{2}}}\right| \leq \frac{3\eta}{a^{\frac{5}{2}}}. \tag{30}$$

Eqs. 29 and 30 and the inequality $\left|\int_0^\infty F(\eta)d\eta\right| \leq \int_0^\infty |F(\eta)|d\eta$, where $F$ is an arbitrary function, give

$$\left|\int_0^\infty \frac{d^3}{d\eta^3}\left(\frac{\sqrt{\eta^2+ia}}{\eta+\sqrt{\eta^2+ia}}\right) e^{\eta(z_1-ix_1)} d\eta\right| \leq \int_0^\infty \frac{6\eta+3a}{a^2}\left(1 + \frac{3\eta^2+2\sqrt{a}\eta}{a}\right) e^{\eta z_1} d\eta, \tag{31}$$

$$\left|\int_0^\infty \frac{d^3}{d\eta^3}\left(\frac{1}{\eta+\sqrt{\eta^2+ia}}\right) e^{\eta(z_1-ix_1)} d\eta\right| \leq \int_0^\infty \frac{3\eta}{a^{\frac{5}{2}}} e^{\eta z_1} d\eta. \tag{32}$$

The right-hand sides of Eqs. 31 and 32 are constant and do not diverge. Replacing $\int_0^\infty \frac{d^3}{d\eta^3}\left(\frac{\sqrt{\eta^2+ia}}{\eta+\sqrt{\eta^2+ia}}\right) e^{\eta(z_1-ix_1)} d\eta$ and $\int_0^\infty \frac{d^3}{d\eta^3}\left(\frac{1}{\eta+\sqrt{\eta^2+ia}}\right) e^{\eta(z_1-ix_1)} d\eta$ with $C_1$ and $D_1$, respectively, the absolute values $|C_1|$ and $|D_1|$ are always less than a constant value. The same applies to $\int_0^\infty \frac{\sqrt{\eta^2+ia}}{\eta+\sqrt{\eta^2+ia}} e^{\eta(z_1+ix_1)} d\eta$ and $\int_0^\infty \frac{e^{\eta(z_1+ix_1)}}{\eta+\sqrt{\eta^2+i}} d\eta$, with $\int_0^\infty \frac{d^3}{d\eta^3}\left(\frac{\sqrt{\eta^2+ia}}{\eta+\sqrt{\eta^2+ia}}\right) e^{\eta(z_1+ix_1)} d\eta$ and $\int_0^\infty \frac{d^3}{d\eta^3}\left(\frac{1}{\eta+\sqrt{\eta^2+ia}}\right) e^{\eta(z_1+ix_1)} d\eta$ being replaced by $C_2$ and $D_2$, respectively. As a result, the electromagnetic fields take the form

$$B_x = -\frac{\mu_0 I}{\pi}\left\{\frac{1}{(z_1-ix_1)} + \frac{1}{\sqrt{ia}(z_1-ix_1)^2} + \frac{2+iaC_1}{ia(z_1-ix_1)^3} + \frac{1}{(z_1+ix_1)} + \frac{1}{\sqrt{ia}(z_1+ix_1)^2} + \frac{2+iaC_2}{ia(z_1+ix_1)^3}\right\}, \quad (33)$$

$$E_y = i\omega\frac{\mu_0 I}{\pi}\left\{\frac{1}{\sqrt{ia}(z_1-ix_1)} + \frac{1}{ia(z_1-ix_1)^2} + \frac{1+(ia)^{\frac{3}{2}}D_1}{(ia)^{\frac{3}{2}}(z_1-ix_1)^3} + \frac{1}{\sqrt{ia}(z_1+ix_1)} + \frac{1}{ia(z_1+ix_1)^2} + \frac{1+(ia)^{\frac{3}{2}}D_2}{(ia)^{\frac{3}{2}}(z_1+ix_1)^3}\right\}. \quad (34)$$

Using Eqs. 33 and 34, $Z_{yx}$ in Eq. 21 is written as

$$Z_{yx} = -i\omega \frac{\left\{\frac{1}{\sqrt{ia}(z_1-ix_1)} + \frac{1}{ia(z_1-ix_1)^2} + \frac{1+(ia)^{\frac{3}{2}}D_1}{(ia)^{\frac{3}{2}}(z_1-ix_1)^3} + \frac{1}{\sqrt{ia}(z_1+ix_1)} + \frac{1}{ia(z_1+ix_1)^2} + \frac{1+(ia)^{\frac{3}{2}}D_2}{(ia)^{\frac{3}{2}}(z_1+ix_1)^3}\right\}}{\left\{\frac{1}{(z_1-ix_1)} + \frac{1}{\sqrt{ia}(z_1-ix_1)^2} + \frac{2+iaC_1}{ia(z_1-ix_1)^3} + \frac{1}{(z_1+ix_1)} + \frac{1}{\sqrt{ia}(z_1+ix_1)^2} + \frac{2+iaC_2}{ia(z_1+ix_1)^3}\right\}}. \quad (35)$$

In the limit $x_1 \to \infty$, the plane-wave assumption is established

$$Z_{yx} = -\frac{i\omega}{\sqrt{ia}} = \frac{-\sqrt{i\omega}}{\sqrt{\mu_0\sigma_1}}, \quad (36)$$

which is also upheld in the limit $z_1 \to -\infty$. This means that if either the horizontal or vertical distance of the localized current is large enough, the plane-wave assumption remains valid. If this condition is not established, the MT responses would be biased by the source field (see Figs. 2–5).

This study focused on the relationship between the MT responses and the vertical/horizontal distances of the source current. Given the numerical simulations (Figs. 2–5 and Figs. S1–S8 in Appendix A) and the above discussions, the source bias diminishes if the altitude of the line source current rises, but becomes large especially when the altitude is within 100–150 km. The altitude distribution of the conductivity in the ionosphere changes temporally/seasonally (Sheng et al., 2014). As shown in Maute and Richmond (2017), although the E layer controls the ionospheric electrical process during daytime, the F layer takes control during nighttime. Therefore, the altitude of the source current can be considered to vary temporally. Assuming the E and F layers are respectively at altitudes 100–150 km and 150–600 km (Sheng et al., 2014), the MT data measured during nighttime possibly indicate a weakened source bias if we use only the line current in the ionosphere for a source. Additionally, when deriving the MT responses using many spectra (i.e., long time-series data), this source bias may be neglected because the electromagnetic fields generated by the localized current are averaged. However, in the estimation of MT impedances from short-term data, the source effect must be considered because the electrical environment in the ionosphere mentioned above changes over a period of several hours to several months. Romano et al. (2014) observed source-dependent temporal changes in the MT responses even at mid-latitudes, where the current system is more likely to be uniform over a wide range than at low latitudes. Therefore, for time-lapse MT, especially at a site above the resistive structure (e.g., 1000 Ωm) and at regions affected by a line source current (e.g., equatorial electrojet), the bias must be considered. Moreover, considering that the range in the biased apparent resistivities, having an initial value of 1000 Ωm, is 550–1,550 Ωm at 200 s (Fig. 5), the MT responses from short-term data must be used with care in the inversion procedure.

**Summary**

The focus in this study was on the bias in the MT responses arising from the variation in altitude and horizontal distances of a localized source current. The numerical examples show that slight changes in distance cause a shift in the MT response especially when the altitude is within 100–150 km, where the E layer exists. These changes in MT responses are possibly seen in real MT data analysis because the vertical distance of the source current varies temporally/seasonally. Therefore, the bias in the MT responses for time-lapse sounding, especially at a region affected by a line source current (e.g., equatorial electrojet) and at a site above the resistive zone, should be evaluated to prevent such changes arising from the source field being regarded as resistivity changes in the subsurface. Moreover, considering that the range in shifts in the MT responses, especially at long periods, depends on source distance, we should treat these responses carefully, for example, by checking the ionospheric environment.

**Appendix A**

Here follow MT responses at periods of 20 s (Figs. S1–S4) and 200 s (Figs. S5–S8) obtained with resistivity values of 100 and 500 Ωm.

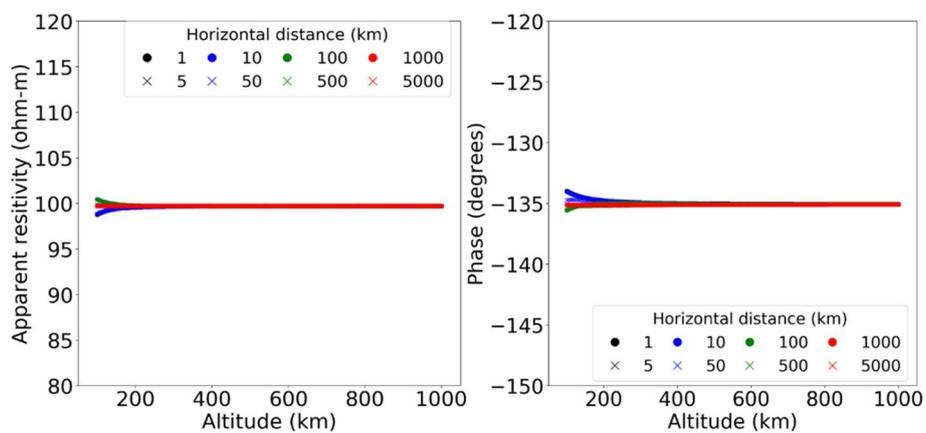

Figure S1. As for Fig. 2 except the subsurface resistivity is set at 100 Ωm.

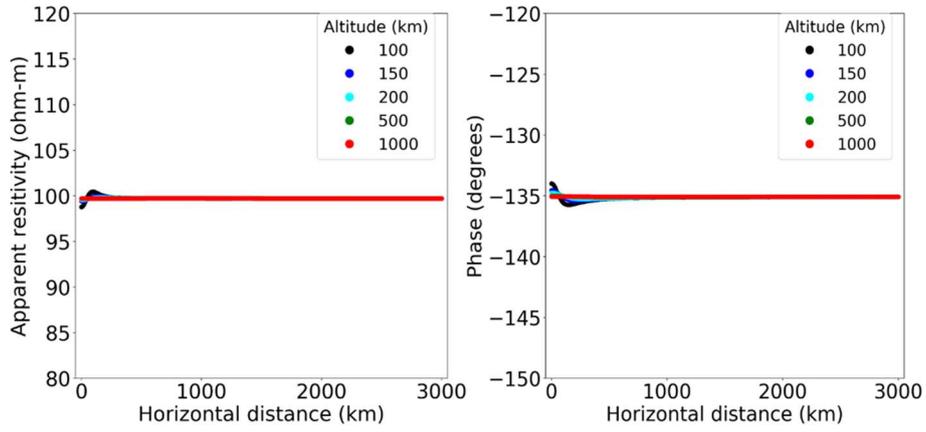

Figure S2. As for Fig. 3 except the subsurface resistivity is set at 100 Ωm.

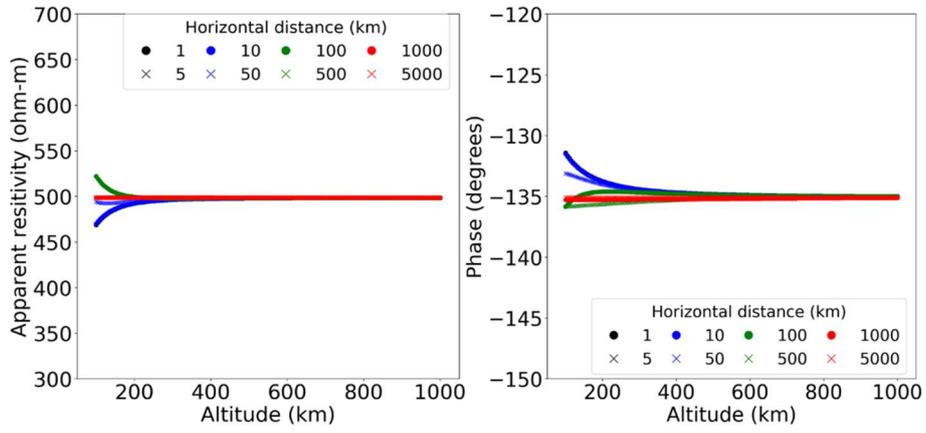

Figure S3. As for Fig. 2 except the subsurface resistivity is set at 500 Ωm.

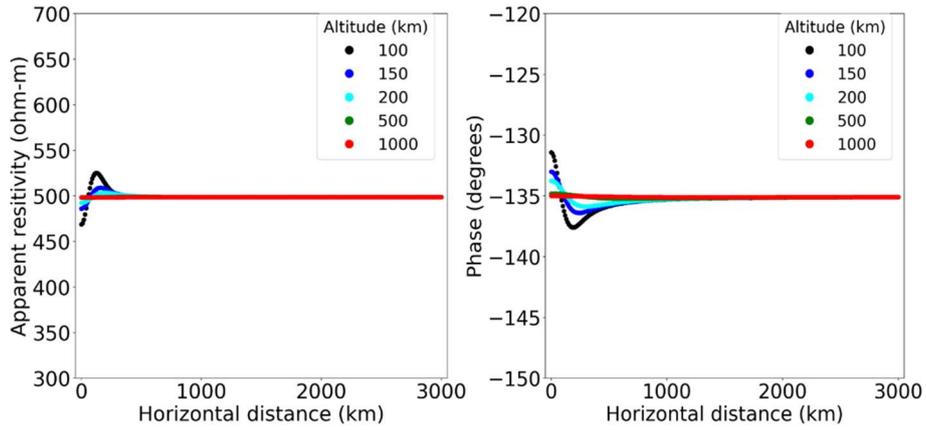

Figure S4. As for Fig. 3 except the subsurface resistivity is set at 500 Ωm.

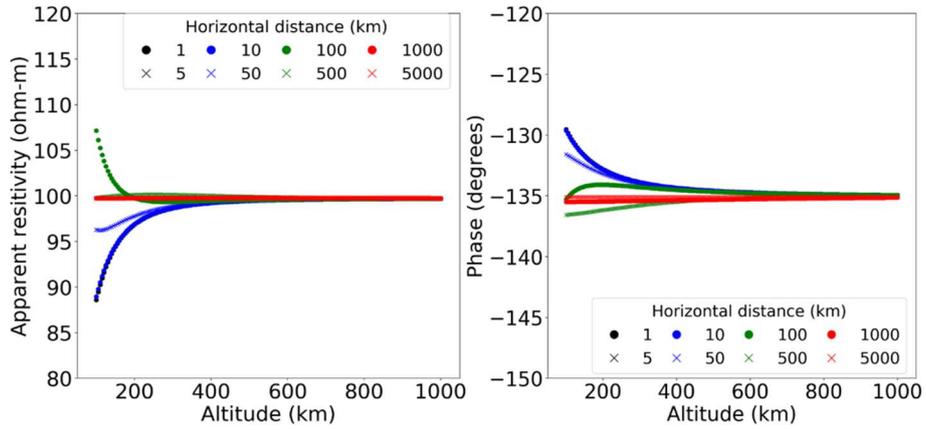

Figure S5. As for Fig. 4 except the subsurface resistivity is set at 100 Ωm.

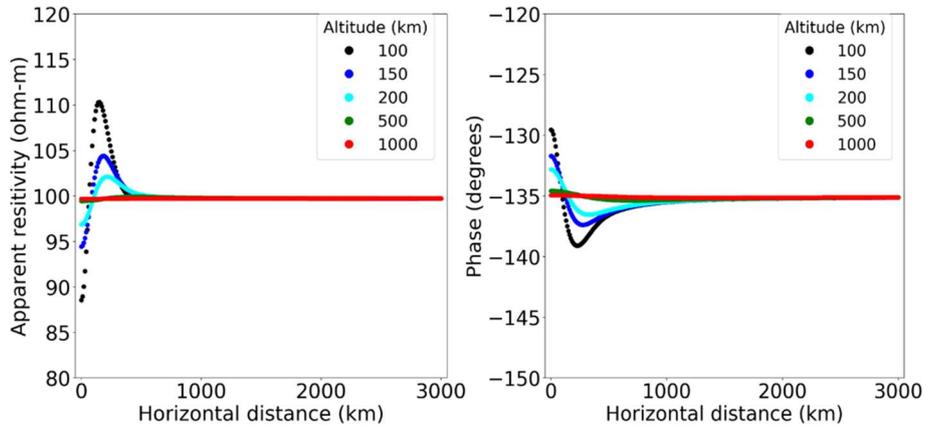

Figure S6. As for Fig. 5 except the subsurface resistivity is set at 100 Ωm.

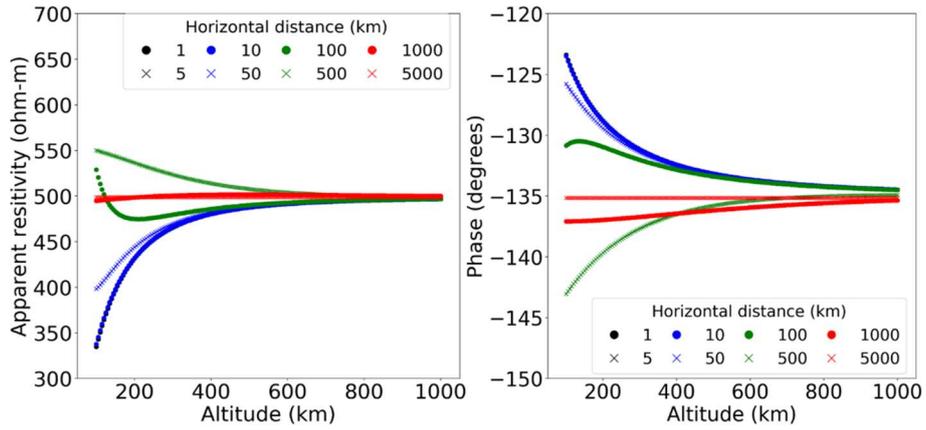

Figure S7. As for Fig. 4 except the subsurface resistivity is set at 500 Ωm.

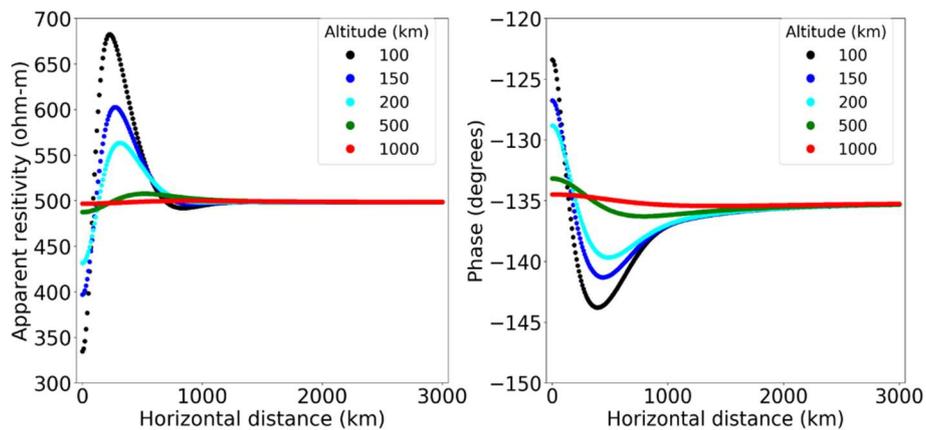

Figure S8. As for Fig. 5 except the subsurface resistivity is set at 500 Ωm.

**Acknowledgements**

The author thanks Dr. Tada-nori Goto, a professor at University of Hyogo, Dr. Katsuaki Koike, a professor at Kyoto University, and Mr. Fumihiko Onoue, a Ph.D. candidate at Scuola Normale Superiore di Pisa, for constructive comments.